%
% the following is to use blackboard bold fonts --
\let\useblackboard=\iftrue
%
% activate this if you don't have them.
\let\useblackboard=\iffalse
%
% You might also need to remove this line.
\newfam\black
\input harvmac.tex
\def\Title#1#2{\rightline{#1}
\ifx\answ\bigans\nopagenumbers\pageno0\vskip1in%
\baselineskip 15pt plus 1pt minus 1pt
\else%\special{papersize=11in,8.5in}%
\def\listrefs{\footatend\vskip 1in\immediate\closeout\rfile\writestoppt
\baselineskip=14pt\centerline{{\bf References}}\bigskip{\frenchspacing%
\parindent=20pt\escapechar=` \input
refs.tmp\vfill\eject}\nonfrenchspacing}
\pageno1\vskip.8in\fi \centerline{\titlefont #2}\vskip .5in}

\ifx\answ\bigans\def\tcbreak#1{}\else\def\tcbreak#1{\cr&{#1}}\fi
\useblackboard
\message{If you do not have msbm (blackboard bold) fonts,}
\message{change the option at the top of the tex file.}
\font\blackboard=msbm10 scaled \magstep1
\font\blackboards=msbm7
\font\blackboardss=msbm5
%\newfam\black
\textfont\black=\blackboard
\scriptfont\black=\blackboards
\scriptscriptfont\black=\blackboardss

\else

\fi
% *************************************
%\draft
%
\def\yboxit#1#2{\vbox{\hrule height #1 \hbox{\vrule width #1
\vbox{#2}\vrule width #1 }\hrule height #1 }}
\def\fillbox#1{\hbox to #1{\vbox to #1{\vfil}\hfil}}
\def\ybox{{\lower 1.3pt \yboxit{0.4pt}{\fillbox{8pt}}\hskip-0.2pt}}

\def\comments#1{}

\def\p{\partial}

\def\I{I}
\def\Ip{I$^\prime$}
\def\II{\relax{I\kern-.10em I}}
\def\IIa{{\II}a}
\def\IIb{{\II}b}
\def\Dslash{\rlap{\hskip0.2em/}D}
\font\cmss=cmss10 \font\cmsss=cmss10 at 7pt
\def\IZ{\relax\ifmmode\mathchoice
{\hbox{\cmss Z\kern-.4em Z}}{\hbox{\cmss Z\kern-.4em Z}}
{\lower.9pt\hbox{\cmsss Z\kern-.4em Z}}
{\lower1.2pt\hbox{\cmsss Z\kern-.4em Z}}\else{\cmss Z\kern-.4em
Z}\fi}
\def\IB{\relax{\rm I\kern-.18em B}}
\def\IC{{\relax\hbox{$\inbar\kern-.3em{\rm C}$}}}
\def\ID{\relax{\rm I\kern-.18em D}}
\def\IE{\relax{\rm I\kern-.18em E}}
\def\IF{\relax{\rm I\kern-.18em F}}
\def\IG{\relax\hbox{$\inbar\kern-.3em{\rm G}$}}
\def\IGa{\relax\hbox{${\rm I}\kern-.18em\Gamma$}}
\def\IH{\relax{\rm I\kern-.18em H}}
\def\II{\relax{\rm I\kern-.18em I}}
\def\IK{\relax{\rm I\kern-.18em K}}
\def\IP{\relax{\rm I\kern-.18em P}}
%\def\IX{\relax{\rm X\kern-.01em X}}
%this doesn't work

\def\IR{\relax{\rm I\kern-.18em R}}
\def\BB{{\bf B}}
\def\BF{\IF}
\def\BR{\IR}
\def\BZ{\IZ}
\def\BR{\IR}
\def\BP{\IP}

\def\RP{\BR\BP}
\Title{ \vbox{\baselineskip12pt\hbox{hep-th/9609071}
\hbox{HUTP-96/A042}
\hbox{RU-96-79}
\hbox{OSU-M-96-22}}}
{\vbox{
\centerline{Small Instantons, del Pezzo Surfaces}
\centerline{and Type \Ip\ theory}}}
\centerline{Michael R. Douglas$^1$, Sheldon Katz$^2$ and Cumrun Vafa$^{1,3}$}
 \medskip
\centerline{$^1$ Department of Physics and Astronomy}
\centerline{Rutgers University }
\centerline{Piscataway, NJ 08855-0849}
 \medskip
\centerline{$^2$ Department of Mathematics}
\centerline{Oklahoma State University}
\centerline{Stillwater, OK 74078, USA}
 \medskip
\centerline{$^3$ Lyman Laboratory of Physics}
\centerline{Harvard University}
\centerline{Cambridge, MA 02138, USA}
 \medskip
\centerline{\tt mrd@physics.rutgers.edu}
\centerline{\tt katz@math.okstate.edu}
\centerline{\tt vafa@string.harvard.edu}
%\centerline{others}
%\smallskip
%\centerline{xxx}
%\centerline{\tt xxx}
\bigskip
\noindent
Samll instantons of exceptional groups arise geometrically
by a collapsing del Pezzo surface in a CY.  We use this
to explain the physics of a 4-brane probe in Type \Ip\ compactification
to 9 dimensions.
\Date{September 1996}
%\draft
%
\def\us#1{\bf{#1}}
\def\nup#1({Nucl.\ Phys.\ $\us {B#1}$\ (}
\def\plt#1({Phys.\ Lett.\ $\us  {B#1}$\ (}
\def\cmp#1({Comm.\ Math.\ Phys.\ $\us  {#1}$\ (}
\def\prp#1({Phys.\ Rep.\ $\us  {#1}$\ (}
\def\prl#1({Phys.\ Rev.\ Lett.\ $\us  {#1}$\ (}
\def\prv#1({Phys.\ Rev.\ $\us  {#1}$\ (}
\def\mpl#1({Mod.\ Phys.\ Let.\ $\us  {A#1}$\ (}
\def\ijmp#1({Int.\ J.\ Mod.\ Phys.\ $\us{A#1}$\ (}
\def\tit#1|{{\it #1},\ }
\lref\dm{M.~R.~Douglas and G.~Moore, ``D-Branes, Quivers, and ALE
Instantons,''
hep-th/9603167.}
\lref\vf{C. Vafa, ``Evidence for F-theory,'' \nup469 (1996) 403;
hep-th/9602022.}
\lref\ganhan{O. Ganor and
A. Hanany, ``Small $E(8)$ Instantons and
Tensionless Non-critical Strings,'' hep-th/9602120.}
\lref\seiwi{N. Seiberg and E. Witten,
``Comments on String Dynamics in Six-Dimensions,''
hep-th/9603003.}
\lref\witmf{E. Witten, ``Transitions in $M$ Theory
and $F$ Theory,'' hep-th/9603150.}
\lref\mv{D. Morrison and C. Vafa, ``Compactifications of $F$ theory on
Calabi-Yau
Threefolds I,'' hep-th/9602114; and ``Compactifications of $F$ theory on
Calabi-Yau
Threefolds II,'' hep-th/9603161.}
\lref\gan{O. Ganor, ``A Test Of The Chiral E8 Current Algebra On A 6D
Non-Critical String,'' hep-th/9607020.}
\lref\ganii{O. Ganor, ``Toroidal Compactification of Heterotic 6D Non-Critical
Strings Down to Four Dimensions,'' hep-th/9608109.}
\lref\kmv{A. Klemm, P. Mayr and C. Vafa,
``BPS States of Exceptional Non-Critical Strings,'' hep-th/9607139.}
\lref\seiberg{N. Seiberg, ``Five Dimensional SUSY Field Theories,
Non-trivial Fixed Points and String Dynamics,'' hep-th/9608111.}
\lref\joerev{S. Chaudhuri, C. Johnson, and J. Polchinski,
``Notes on D-Branes,'' hep-th/9602052.}
\lref\horw{P. Horava and E. Witten, ``Heterotic
and Type I  String Dynamics from Eleven-Dimensions,''
\nup460(1996)506;
``Eleven-dimensional Supergravity on a Manifold with Boundary,''
hep-th/9603142.}
%\lref\mathref{Math ref.}
\lref\witten{E. Witten, ``Small Instantons in String Theory,''
\nup460(1996)541, hep-th/9511030.}
\lref\nsw{K. S. Narain, M. H. Sarmadi and E. Witten,
\nup279(1987)369.}
\lref\ginsparg{P. Ginsparg, \prv{D35}(1987)648.}
\lref\chs{C. Callan, J. Harvey and A. Strominger, \nup367(1991)60.}
\lref\polwit{J.~Polchinski and E.~Witten, ``Evidence for Heterotic-Type I
String Duality,'' hep-th/9510169.}
\lref\bsv{M. Bershadsky, V. Sadov and C. Vafa,
``D-Branes and Topological Field Theories,'' \nup463 (1996) 166.}
\lref\ov{H. Ooguri and C. Vafa, \nup463 (1996) 55; hep-th/9511164.}
\lref\kav{S. Katz and C. Vafa,``Matter From Geometry,''
hep-th/9606086.}
\lref\sen{A.~Sen, ``F-theory and Orientifolds,'' hep-th/9605150.}
\lref\bds{T. Banks, M. R. Douglas and N. Seiberg,
``Probing F-theory with Branes,'' hep-th/9605199.}
\lref\muk{K. Dasgupta and S. Mukhi, ``F-Theory at Constant Coupling,''
hep-th/9606044.}
\lref\nem{J. A. Minahan and D. Nemeschansky, ``An N=2 Superconformal Fixed
Point with $E_6$ Global Symmetry,'' hep-th/9608047.}
\lref\asp{P. S. Aspinwall, \plt371 (1996) 231; hep-th/9511171.}
\lref\dstring{M. Bershadsky, V. Sadov and C. Vafa, \nup463 (1996) 398;
hep-th/9510225.}
\lref\kmp{S. Katz, D. R. Morrison and M. R. Plesser,
``Enhanced Gauge Symmetry in Type II String Theory,'' hep-th/9601108.}
\lref\km{A. Klemm and P. Mayr, ``Strong Coupling Singularities and Nonabelian
Gauge Symmetries in N=2 String Theory,'' hep-th/9601014.}
\lref\witglob{E. Witten, \plt117 (1982) 324.}
\lref\mrd{M. R. Douglas, ``Gauge Fields and D-branes,'' hep-th/9604198.}
\lref\small{M. R. Douglas, D. Kabat, P. Pouliot and S. H. Shenker,
``D-branes and Short Distances in String Theory,'' hep-th/9608024.}
\newsec{Small Instantons, D-branes and del Pezzo surfaces}

It was once thought that the low energy limit of a string theory
was necessarily a field theory.  This may turn out to be true, but a number
of examples have emerged from recent developments in
string theory, for which it is unclear whether a conventional
field theory description
exists.  Whether it does or not,
string theory sheds new light on these phenomena.

An interesting example
is the world-volume theory of a small $E_8$ instanton.
In ten dimensions, this is a six-dimensional theory,
apparently involving a self-dual tensor multiplet and `tensionless strings.'
In lower dimensions, $E_8$ symmetry can be broken, and generalizations
to $E_n$ instantons (for $n<8$) appear.
These theories have been
studied from various viewpoints
\refs{\ganhan,\seiwi,\witmf,\mv,\gan,\kmv,\ganii}.

Recently Seiberg has explained a number of features of the $E_n$
gauge theories in nine dimensional compactification
of type \Ip\ theory,
as properties of the five-dimensional gauge theory of the D$4$-brane
in type \Ip\ theory \seiberg.
The main aim of this note is to explain the observations there
in the framework of \mv.

We understand that related results have been obtained and will appear in
\ref\toap{ D. Morrison, N. Seiberg, to appear.}.

\subsec{Six dimensions}

We start with the M-theory description of small $E_8$ instantons
\ganhan.
Consider the M-theory realization of $E_8\times E_8$ heterotic
strings as compactifications of M-theory on an interval (identified
with $S^1/Z_2$) \horw.  Each $E_8$ lives
on a 9-brane at the boundary of the interval.
M-theory has a 5-brane, whose world-volume carries a self-dual tensor multiplet
of chiral $N=2$ six-dimensional supersymmetry.
When we bring it near the 9-brane,
additional degrees of freedom come down, which respect $N=1$
supersymmetry.
In the limit that it sits at the boundary, it can be identified with
an $E_8$ instanton of zero size.
A new branch of its moduli space appears, on which it remains on the boundary
and fattens up to an $E_8$ instanton of finite size.
By analogy with ordinary gauge theory, the branch where the
five-brane is away from the boundary is a Coulomb branch, with unbroken
tensor gauge symmetry.
The honest instanton is the five-brane on its Higgs branch, in which the
tensor is no longer massless.  Instead, there are 29 new hypermultiplets
describing the moduli of an $E_8$ instanton.

The same process
has an realization in F-theory compactified on a CY 3-fold \seiwi\witmf\mv.
We consider a point on the CY moduli space
where a special 4-cycle shrinks to a point.  This 4-cycle
is a 2 complex dimensional manifold $\BB_n$
known as an $E_n$ del Pezzo surface\foot{As we will note later, the root
lattice of $E_n$ arises geometrically if $n\ge3$.  In addition, ${\bf P}^1
\times {\bf P}^1$ is also a del Pezzo surface.}
 --
we will describe this in more detail below, but it is just ${\bf P}^2$ blown
up at $n\le 8$ points.
It was
conjectured in \mv\ that these correspond to small $E_n$ instantons.

The transition
is described in detail in \mv\ where it was shown that
in the context of F-theory the natural
starting point is $\BB_9$, ${\bf P}^2$ blown up at 9 points
(the 9-th
point being determined by the condition that all points lie
on the intersection of two cubics in ${\bf P}^2$).  If we have
a $\BB_k$ shrinking to a point then we have to get rid of
$9-k$ 2-cycles first.  This can be done by flops, where
one shrinks ${\bf P}^1$ in the del Pezzo to zero size and one
grows another
${\bf P}^1$ in the CY 3-fold which does not lie on the del Pezzo,
going from $\BB_k$ to $\BB_{k-1}$.

In the context of F-theory,
one is looking at elliptically fibered CY 3-folds with zero size
fiber \vf.  In this limit,
all the $\BB_k$ transitions get mapped to the $\BB_8$ transition,
and moreover the one flop needed to go from $\BB_9$ to $\BB_8$
gets identified with the transition point itself.

\subsec{Five dimensions}

Now we compactify an additional circle.
M-theory reduces to type \IIa\ string theory, now
compactified on the $S^1/Z_2$ orientifold, with 16 D8-branes
\refs{\polwit,\joerev}.
This is usually referred to as type \Ip\ theory as it
is T-dual to type I string theory on $S^1$.
The $SO(32)$ Wilson lines on this circle translate directly
into the positions of the D8-branes, and the vacuum with unbroken
$SO(32)$ has all $16$ D8-branes at a single boundary.

The configuration with $E_8\times E_8$ gauge symmetry
has 7 D8-branes at each boundary, and another D8-brane
at a distance $\phi_0$ from each boundary.  The total
length of the interval we identify with type \Ip\ radius
$R$.
The inverse coupling
constant at distance $\phi>0$ from the boundary behaves as $1/g^2=\phi$
for $\phi \le\phi_0$.
For $\phi >\phi_0$, the coupling stops running
and is given by $1/g^2=\phi_0$ until it reaches the D8-brane
at $R-\phi_0$.

To make contact with M-theory, note that in
the strong coupling limit $g\rightarrow \infty$ we have
$\phi_0\rightarrow 0$, and the one D8-brane
meets the 7 D8-branes on the boundary.  We can view the
finite coupling configuration as a `quantum splitting' of the 9-brane of
M-theory.

The compactification reduces the $5$-brane to the D$4$-brane of the
type I' string.  This is the T-dual of the D$5$-brane of type I
string theory, and as such has an $Sp(1)=SU(2)$ world-volume gauge theory.
Its distance from the boundary $\phi$ is the T-dual of an $Sp(1)$ Wilson
line, and thus $\phi\ne 0$ breaks the $Sp(1)$ to $U(1)$.
Thus the Coulomb branch of the small instanton theory is the conventional
Coulomb branch of the D$4$-brane \ganhan.

Now, as the D4-brane moves towards the boundary, it will first hit
the D8-brane which splits from the boundary.
At this point,
an open string stretched between the 4-brane and the D8-brane
becomes massless.  This is the first new physics visible in the
compactification to five dimensions.

Is this in accord with the F-theory description?  This case has
been analyzed in some detail in \refs{\mv,\kmv}.
The compactification of F-theory on a circle of radius $R$,
using the duality chain in \vf, is equivalent to
M-theory on the same elliptically fibered CY 3-fold, but with
an elliptic fiber of size $1/R$.
Now, the $E_8$
del Pezzo transition as described above is a two step procedure,
where we first undergo a flop from $\BB_9$ to $\BB_8$, and then
shrink $\BB_8$ to zero size.  At the first
flop, the mass of the hypermultiplet obtained by wrapping the
M-theory membrane
around the $S^2$ vanishes \witmf.

The two descriptions are in perfect accord!
In fact, this is a particular case of the equivalence
(using T/S dualities) of the two descriptions of massless hypermultiplets,
wrapped D-branes and open string stretched between D-branes, found
in \refs{\bsv,\ov}.

This can also be extended to transitions involving higher del Pezzos.
If we split off extra D8-branes from the boundary,
then as the position of the D4-brane passes through each
one, it corresponds to a flop in the Calabi-Yau.
Let $0\leq \phi_i$ for $i=1,...,k-1$
denote the position of $k-1$ D8-branes, ordered
so that $\phi_i \leq \phi_j$ for $i<j$.

Then the gauge coupling of the 4-brane can be derived either
from the dilaton background produced by the 8-branes \polwit,
or from the quantum dynamics of the 4-brane field theory \seiberg.
The latter description adds to the bare coupling the
one loop gauge coupling renormalization produced by the $SU(2)$
gauge multiplet and by
$SU(2)$ doublet hypermultiplet matter from
strings stretched between D$4$-brane and the D$8$-branes.
These have masses $|\phi\pm\phi_i|$, and in five-dimensional gauge
theory each will produce a linear correction to the gauge coupling.
The result is
\eqn\gaugec{\eqalign{
{1\over g^2} &={1\over g_0^2}+8\phi \qquad \phi < \phi_1 \cr
&={1\over g_0^2}+8\phi_1 +7(\phi -\phi_1)\qquad \phi_1 < \phi <\phi_2 \cr
& .... \cr
&={1\over g_0^2}+8\phi_1 +\sum_{n=2}^{k-1} (9-n)(\phi_n
-\phi_{n-1})+
(9-k)(\phi -\phi_{k-1})\qquad \phi_{k-1}<\phi
}}
where ${1\over g_0^2}$ is the coupling at the boundary.
The $\phi_i$ are identified
with the masses for the $k-1$ doublets of $SU(2)$.
 The limit
of enhanced $E_k$ symmetry is ${1\over g^2}|_{\phi=0}=0$
and $ \phi_i=0$ for $i=0,...,k-1$.
 In other words the zero size
$E_k$ instanton would correspond to $SU(2)$ at infinite
coupling coupled to $k-1$ massless doublets.

We would like to see this detailed description of the gauge
theory in the del Pezzo set up.   In order to do this we will
discuss some mathematical facts about the del Pezzos in the next
section.  The main result is that ${\bf P}^2$ blown up at
$k$ points can be represented as a fibration over ${\bf P}^1$ with
the fibers being generically ${\bf P}^1$ and where at $k-1$ points
over ${\bf P}^1$ we have two ${\bf P}^1$'s intersecting at one point;
in other words over $k-1$ points we will have the geometry of a
resolved $A_2$ singularity.  We will show these facts in the
next section, and then use them to recover the gauge theory
description given above.

\newsec{del Pezzos as fibered spaces over ${\bf P}^1$}

\def\p{{\bf P}}

Recall that ${\bf B}_k$ del Pezzo corresponds to $\p^2$ blown
up at $k$ points.  For the case of $k=1$ there are two
del Pezzos, one corresponding to $\p^2$ blown up at $1$ point
which we denote by
${\bf B}_1$ and the other corresponding to $\p^1 \times \p^1$.
 We show that for $k\geq 1$ this space can
be given a description in terms of a fiber space over $\p^1$.

The construction we will give can be described iteratively.
We will start with ${\bf B}_1$ which is obtained
 by blowing up a point $p$ of $\p^2$.  This is the Hirzebruch
surface ${\bf F}_1$ which is well known to admit a ${\bf P}^1$ fibration
description.  This fibration can be described concretely as follows:
We  pick a line $\ell\subset\p^2$ (recall the algebraic geometry
terminology: a line is the same as a $\p^1$)
not containing $p$.  The fibration $\pi:F_1\to \ell\simeq\p^1$ is defined
by sending $q\in F_1$ to the point of intersection of the line
$pq$ with $\ell$.  The fibers of $\pi$ are identified
with the lines of $\p^2$ passing through the $p$.
Note that this is well defined, even when $q$ lies on
the exceptional divisor of the blown up point $p$.  This is precisely because
we have blown up the point $p$ which basically means we have replaced
it by a sphere worth of points.  This can also be thought of as the point
$p$ together with a tangent vector at $p$.  The tangent vector can
be used to define the line corresponding to the fiber which again
intersects the base $\ell$ at some point.  Thus we have described
${\bf B}^1$ as a space whose base is ${\bf P}^1$ and whose fiber
is also $\p^1$.

We can easily generalize this to ${\bf B}_k$.
Now pick $k$ points $p_1,\ldots,p_k$ and blow up all these points.
We apply the above construction with
$p=p_1$.  The only new thing to consider is what happens when
$q$ lies on the points $p_i$.  Clearly this gets mapped to the
point on $\ell$ where the line $p_1p_i$ intersects it. Let us
call it $q_i$. However
if the point $q$ lies on the exceptional divisor of the blowup
$p_i$ that will also map to the intersection
point of  $p_1p_i$ with $\ell$ which is $q_i$.  Thus the fiber
above $q_i$ consists of two $\p^1$'s, one corresponding to the
line $p_1p_i$ and the other coresponding to the exceptional
divisor of the blowup at $p_i$.  Note that these two $\p^1$'s
intersect at one point.
  Generically $q_i$ are distinct
points on the base.  Thus what we have shown is that ${\bf B}_k$
can be viewed as a fiber space over $\p^1$ where the generic
fiber is $\p^1$, but over $k-1$ points the fiber has two
$\p^1$'s intersecting at a point, i.e. the geometry of a resolved
$A_2$ singularity.

It is important for the application we have in mind to
describe the K\"ahler classes explicitly.  The
dimension of $H^{1,1}({\bf B}_k)$ is $k+1$.  It is convenient
to characterize the dual cycles as follows.  Let us denote the class of
the base of the above fibration by $[B]$ and the class for the generic
fiber by $[A_1]$.  Over the $q_i$ exceptional points on the base
we have $A_2$ resolved singularity which has two classes associated
with it for each of the blow up modes for the $\p^1$'s.
Let us denote these two classes by $[A_2^{i,1}]$ and
$[A_2^{i,2}]$.  These classes are not independent of $[A_1]$.
In fact for each $i$ we have
\eqn\rel{[A_1]=[A_2^{i,1}]+[A_2^{i,2}]}
To see this note that in the neighborhood of each
$q_i$ we have a breaking of $A_2$ to $A_1$ \kav:
$SU(3)\rightarrow SU(2)\times U(1)$.  If we identify the two simple
roots of $SU(3)$ by $e_1,e_2$, then the simple
 root of $SU(2)$ can be identified with $e_1+e_2$ which gives the
relation above.  Thus from two $\p^1$'s over each $q_i$ we get
only one new class which we will define as
\eqn\cla{[A_2^i]=[A_2^{i,1}]-[A_2^{i,2}]}
We have thus given a basis for the $H_2({\bf B}_k)$ consisting
of one class of base $[B]$ one class of the generic fiber $[A_1]$
and $k-1$ classes $[A_2^i]$
of the difference of fiber ${\bf P}^1$'s over $q_i$.
Let $\omega$ denote the Kahler form on the Calabi-Yau pulled
back to the del Pezzo.  We can parametrize the Kahler class
by $k+1$ real numbers which we denote by
\eqn\kcla{\phi =\omega ([A_1]),\qquad \phi_i=\omega([A_2^i])\qquad
k_B=\omega([B])}
With no loss of generality we can take $\phi_i$ to be
postive (if necessary by redefining the order
of the difference of the classes).

Note that $\phi_i<\phi$.  This is because $\omega([A_2^{i,1}])$
and $\omega([A_2^{i,2}]$ are positive numbers whose sum is $\phi$
and difference (in absolute value) is $\phi_i$.  Moreover
precisely when $\phi_i=\phi$ one of the $\p^1$'s
$[A_2^{i,2}]$ has shrunk to zero size.  At this point the
fiber over $q_i$ is a single ${\bf P}^1$.
That is a point in moduli where we can do a flop in the
Calabi-Yau where after shrinking this ${\bf P}^1$
another $\p^1$ in the Calabi-Yau, not contained in the del Pezzo
grows.  We continue to call the corresponding Kahler class
in the Calabi-Yau\foot{Here we are assuming all the classes
in the del Pezzo will be realized non-trivially in the Calabi-Yau.
Otherwise there will be some restrictions on the values of the
$\phi_i,\phi ,k_B$.}
 by $\phi_i$ even for $\phi_i>\phi$.  In this process
the del Pezzo ${\bf B}_k$ has been replaced by ${\bf B}_{k-1}$.
As will be described in more detail in the next subsection, in doing
a flop at each stage we have two choices which are equivalent
except when we get to ${\bf B}_2$.  At this point the
two choices for the flop lead to the two different del Pezzos
$\p^1 \times \p^1$ and ${\bf B}_1$.  Beyond this point only ${\bf B}_1$
can undergo a further flop to $\p^2$.  We cannot do a further flop
on $\p^1 \times \p^1$.

\subsec{More Mathematical Facts about del Pezzos}

It would be interesting to see how unique is our choice
of describing del Pezzos as spaces fibered over ${\bf P}^1$.
We will be examining fibrations $\pi:{\bf B}_n\to {\bf P}^1$ whose general
fiber is ${\bf P}^1$.  Such fibrations are characterized completely by
the cohomology class of the fiber $f$.  The cohomology of ${\bf B}_n$ is
well known, and has as a basis the class $\ell$ of a line in ${\bf P}^2$,
and the exceptional divisors $E_i$ of the blown up points.
The canonical class of ${\bf B}_n$ is $K_{{\bf B}_n}=-3\ell+\sum E_i$, and the
first chern class of ${\bf B}_n$ is given by $c_1=-K_{{\bf B}_n}=3\ell-\sum
E_i$.  The
characterizing property of del Pezzo surfaces says that $c_1$ is ample,
so has positive intersection with every effective curve on ${\bf B}_n$.

If $f$ is the cohomology class of the fiber of $\pi$, then we have $f^2=0$,
and $f\cdot c_1=2$ by the adjunction formula.  An example of such a class
is $f=\ell-E_1$.  The fibers of the corresponding projection map are
lines through the origin as described above.

We now review the Weyl group $W$ of ${\bf B}_n$.  A {\it root\/} is a
cohomology class $v$ with $v^2=-2$ and $v\cdot c_1=0$.  To each root
corresponds a reflection $\rho_v$ on the cohomology

$$\rho_v(D)=D+(D\cdot v)v.$$

Note that $(\rho_v(D))\cdot (\rho_v(D'))=D\cdot D'$ and
$(\rho_v(D))\cdot c_1=0$.  This implies that each
$\rho_v$ preserves the orthogonal complement $V$ of $c_1$, which together
with the set of all roots form a root system.  We let $W=W({\bf B}_n)$ be the
Weyl group of this root system.  Note that $W$ acts as a
group of isometries of the cohomology of ${\bf B}_n$.

We have the following table.

$$\matrix{
n &\hbox{roots} & \hbox{root\ system}\cr
1 & none  & none\cr
2 & E_i-E_j & A_1\cr
3\le n\le 8 & \hbox{generated\ by} \{E_i-E_j,\ell-E_i-E_j-E_k\}& E_n\cr
}$$

In the above, the root systems $E_3,E_4,E_5$ respectively denote
$A_2\times A_1,A_4,D_5$.

Recall that a $-1$ curve is a curve $C$ with $C^2=-1$ and $C\cdot c_1=1$.
These curves are well known.  They are of one of the following types.

$$\matrix{
E_i & \cr
\ell-E_i-E_j & n\ge2\cr
2\ell-\sum_{k=1}^5E_{i_k} & n\ge5\cr
3\ell-2E_i-\sum_{k=1}^6E_{i_k}& n\ge7\cr
4\ell-2\sum_{k=1}^3E_{i_k}-\sum_{m=1}^5E_{j_m}& n=8\cr
5\ell-2\sum_{k=1}^6E_{i_k}-\sum_{m=1}^2E_{j_m}& n=8\cr
6\ell-3E_1-2\sum_{k=1}^7E_{i_k}&n=8
}$$

We now show that any fibration with general fiber ${\bf P}^1$
can be brought to this form by an application of an element of $W$.
That is, we will show that there exists
a $w\in W$ such that $w\cdot f=\ell-E_1$.

Since the Euler characteristic of ${\bf B}_n$ is $n+3$, and ${\bf P}^1$
fibrations
over ${\bf P}^1$ have Euler characteristic~4, we see that not all fibers
can be ${\bf P}^1$ (except in the case $n=1$).  Since $c_1$ is an ample
class, we see that if a reducible fiber $f$ splits up into two pieces
$f=f_1+f_2$, we necessarily have
$f_i\cdot c_1=1$ for $i=1,2$.  Since $f_i$ must be rational,
the adjunction formula gives $f_i^2=-1$.  Also, a fiber cannot be a multiple
fiber, since it would necessarily have multiplicity~2, $f=2g$ with
$g\cdot c_1=0$; but this contradicts the adjunction formula.

If $f=f_1+f_2$ as above, we have $f_i\cdot f=0$, from which we compute
$f_1\cdot f_2=1$.  So the only fibers which are not ${\bf P}^1$s are
a pair of intersecting ${\bf P}^1$s forming an $A_2$ configuration.
Calculation of the Euler characteristic of ${\bf B}_n$ from this geometry
shows that there must be $n-1$ singular fibers of type $A_2$.

We now classify these fibrations.  If $n=1$, then it is easy to see from
$f^2=0$ and $f\cdot c_1=2$ that the only such fibration has $f=\ell-E_1$.
If $n=2$, we similarly see that the only fibrations have $f=\ell-E_1$
and $f=\ell-E_2$, and these are permuted by the reflection in the root
$v=E_1-E_2$.
If $n>2$, then there is a singular fiber consisting of two intersecting
$-1$ curves.  It is well known that we can use the Weyl group to bring
$f_1$ to $E_2$.  We will show that we can then choose an element of the
Weyl group which fixes $E_2$ and brings $f_2$ to $\ell-E_1-E_2$; this
brings $f$ to $\ell-E_1$ as claimed.

{}From the classification of $-1$ curves described above,
the only ones that meet $E_2$ once are

$$\ell-E_2-E_i,$$
$$2\ell-E_2-\sum_{k=1}^4 E_{i_k},\qquad (n\ge5),$$
$$3\ell-E_2-\sum_{k=1}^5E_{i_k}-2E_i,\qquad (n\ge7),$$
$$4\ell-E_2-\sum_{k=1}^4E_{i_k}-2\sum_{j=1}^3E_{m_j},\qquad (n=8),$$
$$5\ell-E_2-E_i-2\sum_{k=1}^6E_{i_k},\qquad (n=8),$$
where in each line, any fixed exceptional divisor appears at most once.

It is straightforward
in each of the cases above to find a series of reflections which
achieves the desired result.
For a class of the first type listed above, we can reflect in the root
$E_1-E_i$.
In fact, each of the classes in the above list can be brought to the
form of a class listed immediately above it by a simple reflection.

We thus conclude that the description of the del Pezzos
as fiber spaces over ${\bf P}^1$ is unique up to the action
of the Weyl group.

As an example of the use of the Weyl group, suppose we flop the exceptional
curve $\ell-E_1-E_2$.  This replaces $\BB_n$ by $\BB_{n-1}$,\foot{This is
only true if $n\ge3$; if $n=2$, then we get ${\bf P}^1\times{\bf P}^1$.}
and preserves the fibration with fiber $\ell-E_1$.  However, due to the
blowdown, $E_1$ is no longer an exceptional curve; we have $E_1^2=0$ in the
blown down surface.  To make contact with our standard description of the
fibration, we choose a root, say $v=\ell-E_1-E_2-E_n$, which has been chosen
so that $\rho_v(E_n)=\ell-E_1-E_2$.  The exceptional curves
$E_i'=\rho_v(E_i),\ 1\le i\le n-1$ are $n-1$ disjoint exceptional curves which
are disjoint from $\ell-E_1-E_2$, hence remain exceptional after the flop.
These $n-1$ curves may be identified with exceptional divisors of $\BB_{n-1}$.
Putting $\ell'=\rho_v(\ell)=2\ell-E_1-E_2-E_n$, we see that the fibration
has fiber $\ell'-E_1'=\ell-E_1$, and we may continue our analysis through
these flop transitions if desired.

\subsec{Generalized del Pezzos}

It is known that it is possible for singular surfaces $S$ to be contained
in smooth Calabi-Yau threefolds and be contractible to a point.  Conditions
for these ``generalized del Pezzo surfaces'' are given in \ref\reid
{M. Reid, G\'eom\'etrie Alg\'ebrique Angers, (A.~Beauville, ed.) (1980)
{\bf 273}.}.
%There are minor technical
%complications if $k<3$, so for simplicity we restrict our attention
%to $k\ge3$.
We list a few of these conditions.

\medskip
\noindent 1. $S$ is Gorenstein (roughly speaking, the singularities of
$S$ are mild enough so that the canonical bundle of the smooth part of
$S$ extends across the singularities to a bundle $\omega_S$ on all of $S$).

\noindent 2. $\omega_S^*$ is ample.

\noindent 3. The reduced, irreducible components of $S$ are surfaces of degree
$a-1$ or $a$ in ${\bf P}^a$, and in particular are either rational or elliptic
ruled surfaces.

\medskip
If $S$ is smooth, then the only possibilities are $S=B_n (0\le n\le 8)$ or
$S={\bf P}^1\times {\bf P}^1$.
%An invariant $k$ can be attached to $S$.  It is most simply described by
%contracting $S$ to a singular point.  Then (if $k\ge3$)
Here we give three examples of generalized del Pezzos which are not smooth
and illustrate the above conditions.

\medskip
\noindent
{\bf Example 1.} Let $C$ be a plane cubic curve, for example
$x^3+y^3+z^3=0$.  We now view the same equation as an equation in ${\bf P}^3$
with an extra coordinate $w$.  The resulting surface $S$ has a singularity
at $(1,0,0,0)$, and $\omega_S^*$ is just the
restriction of the ample class ${\cal O}(1)$ from ${\bf P}^3$ to $S$.
This is an example of an elliptic ruled surface.

This $S$ can be put inside a smooth Calabi-Yau.  For example, the equation
$$x^3+y^3+z^3+f_4(w,x,y,z)+f_5(w,x,y,z)=0$$
is the affine form of the equation of a quintic in ${\bf P}^4$, where $f_i$
denotes a general polynomial of degree~$i$.  It is singular at $(1,0,0,0)$.
We blow up the singularity by replacing $(x,y,z)$ by $(xw,yw,zw)$ and obtain
$$x^3+y^3+z^3+wf_4(1,x,y,z)+w^2f_5(1,x,y,z)=0.$$
This is now smooth due to the presence of a linear term from $wf_4$.
The exceptional divisor is defined by $w=0$, and is identified with the
surface $S$ just described.

\noindent
{\bf Example 2.}  Here we take $S\subset {\bf P}^3$ to be a general cubic
hypersurface with an $A_1$ singularity.  Again, $\omega_S^*$ is just the
restriction of the ample class ${\cal O}(1)$ from ${\bf P}^3$ to $S$.
This $S$ can be put inside the blowup of a quintic in a similar way,
replacing $x^3+y^3+z^3$ above by $w(xy+z^2)+f_3(w,x,y,z)$, with $f_3$
a general cubic.

This example is interesting in that it admits a ${\bf P}^1$ fibration.
One way to see this is that it is a degenerate version
$\tilde{B_6}$ of $B_6$.
The 6~points of ${\bf P}^2$ to be blown up are chosen to lie on a conic $C$,
which therefore is in the class $2\ell-\sum E_i$ on $\tilde{B_6}$.
Then $c_1$ is no longer ample since $c_1\cdot C=0$.  Since $C^2=-2$, shrinking
$C$ yields an $A_1$ singularity on a surface $S$.  It can be shown that any
cubic hypersurface with an $A_1$ singularity is of this form, with $c_1$
corresponding to the hyperplane class \ref\gh{Griffiths and Harris, {\it
Principles of Algebraic Geometry}, Wiley-Interscience 1978.}.

We need to find a fibration $\pi:\tilde{B_6}\to{\bf P}^1$
such that the conic $C$
is contained in the fiber; this guarantees that after collapsing
$C$ to a point, we still have a fibration $S\to {\bf P}^1$.
This can be achieved by taking $f=2\ell-E_1-E_2-E_3-E_4$ (which arises
from reflection of our standard fiber by the root $v=\ell-E_2-E_3-E_4$).
{}From the Weyl group action, we might have expected $5=6-1$ fibers of type
$A_2$ as before.  But something interesting happens.  We expect $A_2$
fibers to arise by applying the Weyl group to $(E_5)+(\ell-E_1-E_5)$ and
$(E_6)+(\ell-E_1-E_6)$.  But we have
$$\matrix{ \rho_v(E_5)&=&E_5\cr
\rho_v(E_6)&=&E_6\cr
\rho_v(\ell-E_1-E_5)&=&2\ell-E_1-E_2-E_3-E_4-E_5&=&C+E_6\cr
\rho_v(\ell-E_1-E_6)&=&2\ell-E_1-E_2-E_3-E_4-E_6&=&C+E_5
}$$

This says that the expected $A_2$ fibers  $\rho_v(E_5)+\rho_v(C+E_6)$
and $\rho_v(E_6)+\rho_v(\ell-E_1-E_6)$ coincide on $\tilde{B_6}$;
they are each $C+E_5+E_6$, an $A_3$.  \foot{Note that $\tilde{B_n}$ is
an auxiliary space in the analysis, and is not contained in a Calabi-Yau.
If we had an $A_n=C_1+\ldots C_n$ fiber in a generalized del Pezzo inside a
Calabi-Yau,
the adjunction formula gives $(C_1+\ldots +C_n)\cdot c_1(\omega_S^*)=2$; but
this is only possible for an ample class if $n\le 2$.}
The curve $C$ lies in the middle of the $A_3$, so after contracting, we get
an $A_2$ conisting of $E_5+E_6$ passing through the singular point of $S$.
We reach a situation with $4$ distinct $A_2$ fibers,
one of which having its intersection point precisely passing
through the singular point of $S$.

\noindent
{\bf Example 3.} For an example with more than one component, consider
$S={\bf P}^1\times A_2$.   Here, $S$ has two components $S_1$
and $S_2$,
each isomorphic to ${\bf P}^1\times {\bf P}^1$.  Each component has
second cohomology ${\bf Z}^2$, and we write cohomology classes as integers
$(a,b)$.  A class is ample if and only if $a>0$ and $b>0$.  We represent the
class of the curve $S_1\cap S_2$ as $(0,1)$ (in either $S_1$ or $S_2$).
The $\omega_S^*$ restricts to each $S_i$ as the class $(2,1)$, hence is
ample.

By degenerating ${\bf P}^1$ to $A_2$, we get a degeneration of
${\bf P}^1\times {\bf P}^1$ to $S$.  No flops would be involved in such a
transition.

\newsec{del Pezzo and $SU(2)$ gauge theory}

If we consider type IIA theory or M-theory on $K3$
near an $ADE$ singularity, it is well known that we end up
with $ADE$ gauge symmetry in 6 or 7 dimension respectively.
Moreover it was argued by employing the adiabatic argument
\refs{\asp,\dstring,\kmp,\km}\ that even if we
consider Calabi-Yau 3-fold
compactifications down to 4 or 5 dimensions, whenever
we have a singularity over a surface of $ADE$ type we obtain
$ADE$ gauge symmetry with $N=2$
supersymmetry in 4 dimensions or $N=1$ in 5 dimensions.
Moreover if the genus of the corresponding surface is $g$
we obtain $g$ adjoint hypermultiplet matter in addition
\kmp \km .  If over the base there are additional
singularities we will get extra matter.
Aspects of these were discussed in \ref\kkm{P. Berglund, S. Katz, A. Klemm and
P. Mayr,``New Higgs Transitions between Dual N=2 String Models,''
hep-th/9605154.}\mv\ref\asgr{P.S. Aspinwall and
M. Gross,``The SO(32) Heterotic
String on a K3 Surface,'' hep-th/9605131.}\ref\sixau{
M. Bershadsky, K. Intriligator, S. Kachru, D.R. Morrison, V. Sadov
and C. Vafa, ``Geometric Singularities and Enhanced Gauge Symmetries,''
hep-th/9605200.}\kav .
In particular the analysis of \kav\ shows that if we have
an $A_1$ singularity fiber which at some points gets enhanced
to $A_2$, what survives of the $SU(3)$ in this fibration
is $SU(2)\times U(1)$ and the corresponding adjoint
is a hypermultiplet in the doublet of $SU(2)$ and charged under
the $U(1)$.

Let us now go back to the del Pezzo $\BB_k$ in view of this
description.   Consider the limit in which
the del Pezzo has a large base $[B]$ with small fiber.  In this
limit using the adiabatic arguments mentioned above we immediately
deduce that we have an $SU(2)\times U(1)^{k-1}$
gauge symmetry with $k-1$ doublets of $SU(2)$
 charged under distinct $U(1)$'s.  Moreover going to the Coulomb
branch of the $SU(2)\times U(1)^{k-1}$ theory corresponds
to resolving the generic $A_1$ singularity of the fiber
and resolving the $k-1$ $A_2$ singular fibers.
We thus immediately identify the $\phi$ defined
in the previous section (the size of the $[A_1]$ fiber)
with the Coulomb branch of $SU(2)$ and the $\phi_i$ with
the difference in Kahler class of the
two $\p^1$'s of the fiber over $q_i$, which can
also be identified with the mass parameters
of the $k-1$ doublets (since they are charged under the $U(1)$'s).
We thus have an $SU(2)$ gauge symmetric theory in its Coulomb branch
with $\phi$ as the expectation value of the scalar
together with $k-1$ doublets of mass $\phi_i$.

The bare coupling constant of $SU(2)$ can be easily identified.
Up to an overall rescaling it is proportional to the area
of the base $[B]$ of the fibration.  This simply follows
from the fact that the effective coupling
in the five dimensional theory gets a volume factor
from the internal space.  Thus (up to a convention dependent
normalization) we identify the $SU(2)$ coupling
${1\over g_0^2}$ with
$${1\over g_0^2}=k_B$$
Note that in the limit of $E_d$ small instanton the whole
del Pezzo shrink to zero size and thus in particular
${1\over g^2}=k_B\rightarrow 0$.
One can also explain the running of the gauge coupling constant
with respect to $\phi_i$ in a geometrical way.  The easiest
way to see this is to start from ${\bf B}_9$, where the coupling
constant does not run.  Then each time there is a flop the coupling
runs.  This comes about geometrically as follows \witmf :
There are interactions of the form
$$\int C_{IJK}A^I F^J F^k$$
in the Calabi-Yau compactification of M-theory
\ref\fere{I. Antoniadis, S. Ferrara and T.R. Taylor, ``N=2 Heterotic
Superstring and its Dual Theory in Five Dimensions,'' hep-th/9511108.},
 where $C_{IJK}$
denote the triple intersection of $H_4$ classes and the $A^I$ are the
gauge fields.  Each time the CY undergoes a flop the classical interesection
number in that class changes and gives rise to a new interaction
of the form $ \int A\wedge F \wedge F$ and by supersymmetry
to an interaction of the form $\int \phi F^2$ \seiberg .
In our case, consider the kahler class
${\cal D}$
dual to the del Pezzo ${\bf B}_k$ sitting in the Calabi-Yau.   We have
$$D\cdot D \cdot D=9-k$$
Note that this Kahler class is responsible for shrinking
the del Pezzo and in particular corresponds to changing of $\phi$.
This classical self-intersection thus
induces an interaction of the form $\int (9-k)\phi F^2$ which
is responsible for the running of the coupling constant with $\phi$.
Each time there is a flop the value of $k$ changes and the dependence
of the gauge coupling on $\phi$ also changes accordingly.

We have thus reproduced all the expected gauge theoretic
properties of the small $E_k$ instantons ($k\not =1$)
in the del Pezzo setup.
As mentioned before for the case of $k=1$ case we have two choices of
del Pezzo and there seems to be a puzzle of which one
we get in the probe theory.  We return to this point
in section 5.

\subsec{Physics of generalized del Pezzos}

As discussed before there are cases in the compactification of
M-theory on Calabi-Yau where the 4-cycle that shrink is
not a smooth del Pezzo, but what is called a generalized del Pezzo.
We gave a few examples of it in the previous section.  It would be
interesting to unravel the physics of them as one might
expect that they lead
to new fixed points of quantum field theories in 5 dimensions.
These may lead to new physics
which may not even be seen by a D4-brane probe
in the simple set up we have.  For instance if we consider example
3 in section 2.2, from the above considerations
we expect to have an $SU(3)$ gauge theory fixed point at infinite
coupling with no matter (there is a mathematical no-go theorem
that suggests that $SU(3)$ with fundamental matter cannot lead
to an interesting fixed point at infinite coupling).  The example 2
of section 2.2 seems to correspond to $SU(2)$ with 4 fundamentals
because it corresponds to an $A_1$ fibration over $\p^1$ where
at four points we have $A_2$ fibration.  However as discussed before
one of the $A_2$ fibers meets the base at a singular point, suggesting
something extra happens to one of the fundamentals.  It would be interesting
to unravel the physics of this extra singularity.
The example 1 of section 2.2 does not admit a fibration
description, so there is no non-abelian gauge symmetry description
of this fixed point.  This is similar to the $\p^2$ case.

\newsec{$D_{n}$ instantons versus $E_{n}$ instantons}

Implicit in the D-brane discussion is the fact that the small
$SO(32)$ instanton and the small $E_n$ instanton, after compactification
on $S^1$, are the same object in different regions of moduli space.
The D$4$-brane which was argued to describe a small $E_n$
instanton is simply the T-dual of the small $SO(32)$ instanton,
a D$5$-brane in type \I\ theory.
In particular the gauge symmetry $SU(2)$ and the matter content was
deduced in this way.
%  Some aspects of the physics, for example the
% existence of strings which become light as
Any truly novel physics of small $E_n$ instantons will appear in the limit
of infinite coupling on the D$4$-brane, as described in \seiberg.

This equivalence bears
some analogy to the example of tensionless strings
associated with degenerating two-cycles in type \IIb\ theory.
After compactification on $S^1$, these are the
T-duals of the massless gauge bosons of type \IIa\ theory.
One might conjecture that such a relation between non-critical low energy
string theories and more conventional field theory is general.

Since the relation between $E_n$ and $D_n$ has not been discussed explicitly,
it may be worth following it in more detail.
We first recall how the two heterotic strings are continuously connected
\refs{\nsw, \ginsparg}.
Start with $SO(32)$; in the compactification the gauge symmetry is
$SO(32)\times U(1)^2$.
By turning on a Wilson line, one can break this to a subgroup such
as $SO(14)^2\times U(1)^4$, and by varying the Wilson line continue to
a point with $E_8\times E_8\times U(1)^2$.

Since there is an unbroken $SU(2)$ throughout the process, one can follow
the small instanton of the $SO(32)$ theory, to produce a small instanton
of the $E_8$ theory, which implies that the two objects are the same.
Indeed there is no quantum number in the five-brane solution of
\chs\ available to distinguish them -- the only possible distinction could
be the embedding of $SU(2)$ in the gauge group, and one can check that
this process produces an instanton with the same (minimal) embedding of
$SU(2)$.

This leads to a bit of a paradox, as it implies that all of the strange
physics of the $E_8$ small instanton should be implicit in the rather
tame-looking $SO(32)$ small instanton.
This is essentially the same paradox that appeared in the early discussion
of heterotic -- type \I\ duality.  Duality requires that
the heterotic string world-sheet
physics responsible for gauge symmetry enhancement to $E_8$ be
present in the type \I\ string.  Since it is true for any value of the
heterotic string coupling including strong coupling,
it must be visible at weak type \I\ coupling,
which naively rules out the possibility of solitonic states becoming light.

The resolution was that type \I\ perturbation theory still breaks down,
due to a
failure of the dilaton tadpoles from the disk and $\RP^2$ to cancel in
winding sectors.  This effect becomes important as the
type \I\ compactification radius becomes small, and indeed
the region of heterotic string moduli space
with enhanced gauge symmetry always maps into this regime \polwit.
The resulting physics
is much clearer in the T-dual picture, where the dilaton
becomes strong at a fixed point of the orientifold, but in principle
all of these effects could be translated back to the original type \I\
picture.
For example, the D$0$-branes which became the gauge bosons of $E_8$
gauge symmetry in \joerev, correspond to type \I\ heterotic solitons
wrapped around a small circle.

Thus, the statement is that $D_{n-1}$ gauge symmetry in either theory
can be promoted either to $D_n$ or to $E_n$ by tuning different
parameters.  In the type \Ip\ description, $D_n$ is achieved by
bringing another D$8$-brane to the boundary, while $E_n$ is achieved
by going to infinite coupling on the boundary.
A zero size instanton which is present will gain the
additional massless states of $D_n$ or $E_n$ global symmetry, and
the corresponding additional dimensions on its Higgs branch, thanks to
the same physics.

The case of $D_5 \cong E_5$ symmetry is an interesting illustration.
The theory (at one boundary)
has manifest $SO(8)\times U(1)^2$ gauge symmetry, which can
be enhanced to $SO(10) \times U(1)$ in two ways.
If another $8$-brane is brought to the boundary, stretched open strings
in vector multiplets come down, producing
$45 = 28_0 + 1_0 + 8_{v,1} + 8_{v,-1}$.
On the other hand, by adjusting parameters to take the strong coupling
limit at the boundary, D$0$-branes become massless,
charged under a different $U(1)$,
and with fermion zero modes (from 0--8 strings) in the vector of $SO(8)$.
Quantizing these puts the $0$-branes in spinor representations
of $SO(8)$, to produce
$45 = 28_0 + 1_0 + 8_{s,1} + 8_{c,-1}$.

The global symmetry on the D$4$-brane is also enhanced.
Electrically charged matter $Q_i$ from the 4--8 strings transforms in the
vector of $SO(8)$, and in the $D_5$ case becomes the vector of $SO(10)$.
In the $E_5$ case, since the $4$-brane gauge coupling is strong, it is
only sensible to consider gauge singlet operators, such as the bilinears
$Q_i Q_j$ in the adjoint of $SO(8)$.  These must be joined by
0-brane -- 4-brane bound states to fill out the adjoint of $SO(10)$, and
the simplest way this could work is
if each 0-brane which appeared as a gauge multiplet can also appear
as part of a unique multiplet of 0--4 bound states.
The same story could account for $E_n$ global symmetry, and
it would be quite interesting to check it against precise results for
0--4 bound states.

Additional $4$-brane BPS states are predicted in \refs{\gan,\kmv},
and it might be even more interesting to identify these,
as some of them are singlet under the $E_n$ symmetry, and thus
need not involve D$0$-branes.

The continuous connection between the D$5$-brane  and the
D$4$-brane appears to require
string theory, but this is not incompatible with the idea that the
non-trivial physics of the D$4$-brane and the origin of the $E_n$
massless states has a purely field theoretic explanation.
The D$0$-branes of the string description would be equivalent to
instantons of the five-dimensional gauge theory.

\newsec{$\BF_1$ versus $\BP^1\times \BP^1$ and a discrete $\theta$ angle}

The detailed agreement between the del Pezzo description and the gauge
theory description for every other case leads us to return to the puzzling
example of the surfaces $\BF_1$ and $\BP^1\times \BP^1$, and try to
understand this in gauge theory terms.

Both surfaces can be realized by flopping a $\BP^1$ in $B_2$, so let
us describe the choice which distinguishes them.
In general, we can flop any of the exceptional curves,
and since all of them are related by the Weyl group, the result is
the same.  But for $B_2$, the curves $E_2$ and $\ell -E_1 -E_2$ are
not so related.  Since $\phi$ is the size of the $A_1$ fiber
$\ell -E_1$, and $\phi_1$ is the size of $\ell -E_1 -2E_2$,
the two exceptional curves correspond to states with mass
$(\phi-\phi_1)/2$ and $(\phi+\phi_1)/2$.

Thus, the mathematics appears to be telling us that the definition
of pure $SU(2)$ supersymmetric gauge theory in five dimensions involves
a subtle two-valued choice, which upon adding massless matter
becomes vacuous.  Starting from the theory with one matter multiplet,
we can recover the pure gauge theory by taking its bare mass to infinity,
but we will get different results depending on whether we take it to
positive or negative infinity (relative to the sign of $\phi$; of course
an $SU(2)$ gauge transformation flips the sign of all $\phi_i$ and $\phi$).

This is very reminiscent of the way the $\theta$ angle in $d=4$ gauge
theory becomes vacuous in a theory with massless fermions, and
is affected by the phase of the fermion mass terms.
Thus we ask whether $d=5$ $SU(2)$ gauge theory admits
a $\BZ_2$-valued $\theta$ angle.

Just as the $\theta$ angle in $d=4$ weighs gauge field configurations by
$\exp {i\theta n}$ with $n$ the instanton number in $\pi_3(SU(2))$,
a $\BZ_2$ $\theta$ angle in $d=5$ will exist if
$\pi_4(SU(2)) \cong \BZ_2$.
Indeed this is true, and we identify this $\theta$ angle as the
two-valued choice.

We next need to understand its relation to a fermion mass $m$.
In $d=4$ this was a consequence of the axial anomaly.
To use an argument which generalizes to the discrete case,
it follows from the existence of $2nk$ fermion zero modes in the $n$-instanton
sector, requiring $nk$ insertions of the mass term $m\bar\psi\psi$
for a non-zero amplitude, and weighing this sector by $m^{nk}$.

Now, a known consequence of the fact that $\pi_4(SU(2)) \cong \BZ_2$ is
the global anomaly in $d=4$ $SU(2)$ gauge theory with an odd number
of fermion doublets \witglob.
This is a sign ambiguity in the fermion determinant
$(\det \Dslash)^{1/2}$ under a global gauge
transformation $U$ in the non-trivial class of $\pi_4(SU(2))$.
In \witglob, this ambiguity was exhibited by finding a non-trivial
path in configuration space connecting a gauge field configuration $A^{(0)}$
with the configuration gauge transformed by $U$, and showing that an odd
number of eigenvalues of the Dirac operator will change sign along the
path.
Such a path defines a five-dimensional gauge field in the non-trivial
class of $\pi_4(SU(2))$, and this is equivalent to the statement that the
mod two index for the five dimensional Dirac operator will be odd in
this field configuration.

This fact also implies that in our five dimensional theory, changing the
sign of a single fermion mass will flip the $\BZ_2$ theta angle.
Each $8$-brane and its image comes with a half-hypermultiplet in the $(2,2)$
of $SO(2)\times SU(2)$, corresponding to two Weyl doublets in the $d=4$
theory (we know that the $d=4$ reduction is non-anomalous), and a fermion
determinant $\det (\Dslash+m)$, defined unambiguously by combining the
fermions into a single Dirac fermion.
The operator $\Dslash$ is real antisymmetric and in a sector with
$k$ fermion zero modes, the determinant is a product over eigenvalues
$$
\det (\Dslash+m) = m^k\ \prod_i (\lambda_i+m)(\lambda_i-m),
$$
proving the result.
We thus find that the moduli space of
five dimensional gauge theory mirrors
the moduli space of del Pezzo surfaces in every detail.

Now that we have shown that the distinction between
$\BF_1$ and $\BP^1\times \BP^1$ is a distinction between two different
$SU(2)$ gauge theories, let us discuss the difference between the physics
of the two cases.
The main difference is that whereas $\BP^1\times \BP^1$ can only be shrunk
to zero volume in a single way, $\BF_1$ has two independent parameters.
In particular, it admits a final flop, down to $\BP^2$.
Geometrically the membrane wrapped around the exceptional
divisor becomes massless at this point, which implies that for
${1\over g^2 }=c \phi$ we have a massless hypermultiplet BPS state
with charge $1$ under the $U(1)$.  Beyond
this point the running of the coupling should change, as discussed
before, to $1/g^2 \sim {\rm const.}-9\phi$.  Moreover beyond
this point we do not have any fiber space description which
implies that even at $\phi =0$ we do not have the massless
$W^{\pm}$ bosons.  This suggests that after the last flop
the $W^{\pm}$ become unstable.

   The appearance of a
new massless charged BPS state in the pure $SU(2)$ gauge theory
at finite coupling is somewhat surprising and deserves explanation.
Since the BPS mass formula for this state involves $1/g^2$,
this must be a bound state involving a D$0$-brane (or instanton).
Indeed the total charge is that of a bound state with the massive $SU(2)$
gauge boson, and the positive instanton mass $1/g^2$ is being canceled by
an opposite sign central charge for the W boson.
At this point the $W^{\pm}$ become marginally unstable to decay
into the instanton BPS state and the new hypermultiplet, and the
geometry as well as the running of the coupling
beyond the transition indicate that they become unstable.

The point in $S^1/\BZ_2$ where the bound state becomes massless is
not associated with any feature of the
background, and this might be thought to be a contradiction to the
general `probe' philosophy \refs{\mrd,\dm,\bds,\small}\ that gauge dynamics
on the brane must reproduce the background fields, in this case the dilaton.
Now for the discrete $\theta$ angle
corresponding to $\BP^1\times \BP^1$, the probe philosophy works.
Furthermore, we can take several probes and take some of them through the
orientifold point from $\phi$ to
$-\phi$, thus choosing independent $\theta$ angles
for each.  Thus we would associate the $\BF_1$ dynamics not to
a difference in the background but rather with a choice made on the $4$-brane
which makes it not act as a probe.  One might entertain other interpretations
and perhaps more can be said about this issue.

A comment which may make the result more palatable than it seems at
first is that similar results are implied by the global symmetry in the
$N_f\ge 1$ cases.  The Weyl group of the full enhanced symmetry will
act on the moduli space and spectrum, and exchange points of moduli
in the gauge theory parametrization which do not seem to look
like symmetries of the perturbative spectrum.  As discussed before
the resolution is that under the Weyl group
the perturbative states mix with  non-perturbative solitonic objects
in the field theory language, i.e. BPS instanton D0-branes.
Our present discussion of the flop to $\BF_1$
is just the simplest manifestation of this.

\newsec{Discussion}

We found agreement between two descriptions of the Coulomb branch
for $E_n$ zero size instantons in five dimensions,
one provided by F-theory and M-theory on a Calabi-Yau
containing a degenerating del Pezzo surface \mv,
the other provided by the five-dimensional gauge
theory physics of D$4$-branes in type \Ip\ theory \seiberg.
We also found hints of new five-dimensional fixed point
theories, for which we do not know of a D-brane description,
by considering generalized del Pezzos.

We also discussed some general properties of
the Higgs branch of the gauge theory, describing the true small instantons.
For example, we saw that the $D_n$ and $E_{n+1}$ small instantons
are continuously connected, and the symmetry enhancement of the infinite
coupling limit is associated with additional $0$-brane (or instanton)
states becoming massless.  Their associated fields provide the
additional parameters of $E_{n+1}$ instanton moduli space,
which enter on an equal footing with the $D_n$ parameters.

Even though many aspects of the physics associated
with small $E_n$ instantons can be captured by
gauge theory dynamics, one should keep in mind
that this is at a fixed point at infinite coupling, and thus
the spectrum is not evident from this description.
The underlying dynamics of the theory remains somewhat mysterious.
For example, the geometric description predicts
a tensionless string in five dimensions,
from a 5-brane wrapped around the vanishing 4-cycle \witmf.
This string will be as relevant as all the massless particle
states coming from membranes wrapped around 2-cycles
inside the del Pezzo.
On the other hand, the origin of this string in the gauge theory
is the 't Hooft-Polyakov monopole solution \seiberg.
As $SU(2)$ is restored, it might be thought that
this solution would delocalize and should not
be considered part of the spectrum, but this is at finite coupling.
It is consistent with what we know so far to imagine that
at the non-trivial fixed point, it remains in the theory.

It remains to develop effective ways to discuss the physics
at the fixed point.  In this regard it is tempting to conjecture,
given that ${1\over g^2}\rightarrow 0$ that we are left with
a pure $N=1$ Chern-Simons theory in 5 dimensions.  This suggests that
the relevant aspects of the fixed point may be captured by
a topological supersymmetric Chern-Simons theory in 5 dimensions.

The situation will change when one compactifies
one more dimension.  In this case the formal
scaling arguments suggest that the leading massless
modes are particles \kmv\ and there should be a more conventional
field theory description.
This is in fact supported by the recent study of this case \ganii.
We will be very brief here and only
point out some basic features relevant for further study.
In this case it is natural to consider the F-theory compactification on $K3$
\vf\ and study a 3-brane probe following
\refs{\sen,\bds,\muk,\nem}.
It may appear that we will see a clash between the
del Pezzo description and the 3-brane probe behavior
near an $E_{6,7,8}$ enhanced gauge symmetry point.
We expect
from \muk\ that the corresponding gauge coupling $\tau$ on the probe is
at fixed but finite values (at the fixed
points of $SL(2,{\bf Z})$ elements of order $3,4,6$).  However
the type IIA description on a CY 3-fold with a
zero size del Pezzo, which is what we obtain
after compactification of M-theory on circle, might seem to
lead to infinite coupling as before.

We believe the resolution of this puzzle may be that
for type IIA (unlike the M-theory case), because we
have corrections to the volume of Calabi-Yau due to
worldsheet instantons,\foot{C.V. thanks B. Greene for reminding
him of this in the present context.}\ %
at the transition point the quantum volume of the base
(together with the $B$-field) will not be zero but instead be given by
the fixed point values of $\tau$ predicted by \muk.  This
will be very interesting to study.  In particular the results
of \ganii\ should be derivable by using mirror symmetry
acting on del Pezzo (along the lines of the appendix in \kmv ).

\medskip

We would like to thank B. Crauder, B. Greene,
W. Lerche, D. Morrison, N. Seiberg and N.P. Warner for valuable
discussions.
C.V. would like to thank the hospitality of Rutgers University
where this work was done.

The research of M.R.D. was supported in part by DOE grant DE-FG02-96ER40559,
that of S.K. was supported in part by NSF grant DMS-9311386 and
NSA grant MDA904-96-1-0021,
and that of C.V. was supported in part by NSF grant PHY-92-18167.

\vfill\eject
\listrefs
\end